\documentclass{aa501}
\usepackage{graphicx}
\begin{document}

\def\bril {mJy~beam$^\mathrm{-1}$}
\def\simlt{\mathrel{\rlap{\lower 3pt\hbox{$\sim$}}
        \raise 2.0pt\hbox{$<$}}}
\def\simgt{\mathrel{\rlap{\lower 3pt\hbox{$\sim$}}
        \raise 2.0pt\hbox{$>$}}}

\title{Radio--Optically Selected Clusters of Galaxies}
\subtitle{I. The Radiogalaxy Sample}

\author{A.~Zanichelli \inst{1}
        \and M.~Vigotti \inst{1} 
        \and R.~Scaramella \inst{2}
        \and G.~Grueff \inst{3}
        \and G.~Vettolani \inst{1}
       }

\offprints{A. Zanichelli, \email{azanichelli@ira.bo.cnr.it}}

\institute{Istituto di Radioastronomia -- CNR, Via Gobetti 101, I--40129 
                Bologna, Italy
           \and Osservatorio Astronomico di Roma, via Osservatorio 2, I--00040 
                Monteporzio Catone (RM), Italy
           \and Dipartimento di Fisica, Universit\`a di Bologna, Via Irnerio
                46, I--40127 Bologna, Italy
          }
\date{Received ... / Accepted ...}

\abstract{In order to study the status and the possible evolution of clusters 
of galaxies at intermediate redshifts ($z \sim 0.1 - 0.3$), as well as their
spatial correlation and relationship with the local environment, we built a
sample of candidate groups and clusters of galaxies using radiogalaxies as
tracers of dense environments.
This technique -- complementary to purely optical or X-ray cluster selection
methods -- represents an interesting tool for the selection of clusters in a
wide range of richness, so to make it possible to study the global properties 
of groups and clusters of galaxies, such as their morphological content, 
dynamical status and number density, as well as the effect of the environment 
on the radio emission phenomena.
In this paper we describe the compilation of a catalogue of $\sim 16\,000$
radio sources in the region of the South Galactic Pole extracted from the
publicly available  NRAO VLA Sky Survey maps, and the optical identification
procedure with galaxies brighter than $b_\mathrm{J} = 20.0$ in the EDSGC 
Catalogue.
The radiogalaxy sample, valuable for the study of radio source populations down
to low flux levels, consists of $1288$ identifications and has been used to 
detect candidate groups and clusters associated to NVSS radio sources.
In a companion paper we will discuss the cluster detection method, the cluster
sample as well as first spectroscopic results.
 }

\titlerunning{Radio--Optically Selected Galaxy Clusters. I}
\authorrunning{A. Zanichelli et al.}

\maketitle

\section{Introduction}\label{sec:intro}

One of the major topics in modern cosmology concerns the dynamical status and 
evolution of groups and clusters of galaxies, as well as their abundance and 
spatial distribution, their morphological content and interactions with the 
environment. Groups and clusters of galaxies are indeed the largest, 
gravitationally bound, observable structures, and by studying their properties
and the processes underlying their formation much can be understood about the
global cosmological properties of the universe.

In recent years, significant efforts have been made in searching for clusters 
at high redshifts; nevertheless the general properties and the physical 
processes at work in these large scale structures at moderate $z$ are still 
unclear.
To this aim, it is of fundamental importance to gather cluster samples
representative of different dynamical structures -- from groups to rich
clusters -- in a wide range of redshift and covering large areas of the sky.

First attempts to build wide-area cluster samples, like the ACO/Abell catalogue
(Abell et al. \cite{Abell}), were based on visual inspection of optical plates 
and only recently the first catalogues obtained through objective algorithms 
appeared ( EDCC, Lumsden et al. \cite{Lumsden}; APM, Dalton et al. 
\cite{Dalton}).
The selection based on optical plates, however, limits the redshift range to 
about $z < 0.2$ and suffers from misclassifications due to projections effects 
along the line of sight, resulting on one side in spurious cluster detection 
and, on the other side, in wrong estimates of the cluster richness, that can
affect the reliability of the derived cosmological parameters (van Haarlem 
et al. \cite{van Haarlem}).
Alternatively, cluster samples at higher redshift have been built using a 
matched filter algorithm which makes use of both positional and deep multiband 
CCD photometric data over selected areas of few square degrees (Postman et al. 
\cite{Postman}; Scodeggio et al. \cite{Scodeggio}).

Also, to find candidate clusters at intermediate redshift through color
diagrams alone could bias the selection against clusters with a high fraction 
of blue galaxies, whose presence can be due to the occurrence of the 
Butcher--Oemler effect (Butcher \& Oemler \cite{Butcher}) or to the fact that 
the cluster itself can be in the process of formation.

The X-ray emission properties of the hot intracluster medium have been widely 
used to build distant cluster samples, but this technique suffers from the 
limited sensitivity of wide-area X-ray surveys and from the possibility of 
evolutionary effects (Gioia et al. \cite{Gioia}; Henry et al. \cite{Henry}; 
RDCS, Rosati et al. \cite{Rosati}).

Even more critical is the selection of groups of galaxies: these structures
-- which represent a sort of ``bridge'' between rich clusters and the field -- 
are of major interest for the understanding of galaxy interactions and 
evolutionary processes, but their detection is particularly difficult even at 
moderate redshifts due to their very low density contrast with respect to 
field galaxies distribution.

A different approach -- complementary to purely optical or X-ray cluster
selection methods -- is the use of radiogalaxies as suitable tracers of dense 
environments. In recent studies (Prestage \& Peacock \cite{Prestage}; 
Hill \& Lilly \cite{Hill}; Allington--Smith et al. \cite{Allington--Smith}; 
Zirbel \cite{Zirbel97}; Miller et al. \cite{Miller}) it has been shown that 
Faranoff--Riley I and II radio sources are found in different environments, and
differ in the optical properties of their host galaxies as well.
FRI sources are found on average in rich groups or clusters at any redshift, 
and are associated with elliptical galaxies, with the most powerful FRIs often 
hosted by a cD or double nucleus galaxy (Zirbel \cite{Zirbel96}).
FRII radio sources are typically associated with disturbed ellipticals and 
avoid cD galaxies, and at $z\sim 0.5$ FRIIs are found in a wide range of
environments, including many rich clusters which rarely, if ever, host a FRII
radio source at low redshift (Zirbel \cite{Zirbel96}; 
Hill \& Lilly \cite{Hill}).

Radio selection should not impact on the X-ray or optical properties of the
cluster found in this way, since there is no significant correlation between 
the radio properties of galaxies within a cluster with its $L_\mathrm{X}$ 
(Feigelson et al. \cite{Feigelson}; Burns et al. \cite{Burns}), or with 
richness of the cluster (Zhao et al. \cite{Zhao}; 
Ledlow \& Owen \cite{Ledlow}).
Moreover, since no correlation exists between the properties of group members 
and the radio characteristics of the radiogalaxies, radio-selected groups can 
be used to study the general evolution of galaxies in groups 
(Zirbel \cite{Zirbel97}).

Radiogalaxies can thus be used as tracers of dense environments at any epoch,
and the evolution of galaxy groups and clusters can be studied lessening those 
biases that are the main drawbacks of pure optical or X-ray selected cluster 
samples.

A further point that makes this selection technique interesting is the 
possibility to investigate the effects of the environment on the radio-emission
phenomena. Zirbel (\cite{Zirbel97}) speculates the possibility of two distinct 
scenarios for the fueling of radio emission in FRI and FRII sources.
The difference in the environments of FRII radio sources at low and high 
redshift suggests that the conditions to form such sources have changed with 
epoch, and the characteristics of their optical counterparts are consistent 
with the hypothesis of FRII radio emission being fueled by galaxy encounters. 
For FRI radio sources, it is suggested the possibility of them 
being drawn from different galaxy types, and being triggered by different
mechanisms, depending on their power.
The most powerful FRI sources are typically dominant galaxies and their 
environments seem to be consistent with the possibility of them being cooling 
flow galaxies: in this scenario, the cooling flow itself can provide the fuel 
for the radio source.
The less powerful FRI sources do not always correspond to the first ranked 
galaxy, are not always found in the centre of the potential well, and some
reveal signs of galaxy interactions (see e.g. Baum et al. \cite{Baum}). 
It seems thus unlikely that the less powerful FRIs can be cooling flow 
galaxies, and the radio emission could be triggered by a different mechanism
with respect to more powerful FRIs.

This scenario suggests that the radio source morphology is not only a function 
of the radio power, as suggested by theoretical models 
(Bridle \& Perley \cite{Bridle}; Bicknell \cite{Bicknell84}, 
\cite{Bicknell86}), but depends also on the epoch of observation, that is the
density and evolution of the intracluster medium.
In this sense, the study of radio-selected groups and clusters over a wide 
range in radio power may help in understanding the physics of radio emission
and the relationships between different classes of AGN.

To build such a sample of radio-traced clusters, the new radio surveys NRAO VLA
Sky Survey  (NVSS, Condon et al. \cite{Condon}) and Faint Images of the Radio 
Sky at Twenty--centimeters (FIRST, Becker et al. \cite{Becker}) offer an 
unprecedented possibility to study a wide-area, homogeneous sample of radio 
sources down to very low flux levels, together with a positional accuracy 
suitable for optical identifications.

Recently, Blanton et al. (\cite{Blanton}) looked for moderate to high redshift 
clusters associated with a sample of radio sources from the FIRST survey,
having a bent-double radio morphology. The presence of a distorted radio 
structure may be the consequence of the relative motion of the host galaxy in 
the intracluster medium, or of tidal interactions with other cluster galaxies, 
and thus can be used as an indicator of the presence of a cluster or group 
surrounding the radio source.
From R-band imaging of the field surrounding bent-double radio sources, Blanton
et al. (\cite{Blanton}) selected ten candidates for multislit spectroscopy, and
for eight of them they found evidence of a cluster associated to the 
radiogalaxy, with measured richnesses ranging from Abell class 0 to 2. As FRI 
sources more frequently show a distorted morphology, this sample contains 
mostly FRI radiogalaxies.
Moreover, due to its high resolution, the FIRST survey may resolve out extended
sources, making the FRI/FRII classification difficult.

The lower angular resolution of the NVSS survey makes this survey more suitable
than the FIRST for the detection of extended regions of low surface brightness.
We used the publicly available radio maps in the NVSS to build a sample of 
radio-optically selected clusters associated to FRI and FRII radio sources over
a wide area in the sky. In this paper we describe the compilation of a radio 
source catalogue and the optical identification procedure with galaxies in
the EDSGC Catalogue (Nichol et al. \cite{Nichol}) that led to the compilation 
of the radiogalaxy sample.
In a companion paper (Zanichelli et al. \cite{Zanichelli}, Paper II) we will 
present the cluster selection method and the sample of candidate clusters we 
have obtained, as well as first spectroscopic results.

This paper is structured as follows: in Sect.~\ref{sec:radiodata} we give a 
description of the characteristics of NVSS radio data and discuss the need to 
compile a radio source catalogue in alternative to the NVSS publicly available 
one. The extraction of the radio source catalogue is presented in 
Sect.~\ref{sec:catextraction}, together with a discussion on the classification
of double radio sources. The radio source catalogue and its properties are
discussed in Sects.~\ref{sec:radiocat} and ~\ref{sec:testcat}. In 
Sect.~\ref{sec:optids} and following the optical identification procedure and 
the obtained radiogalaxy sample are described.

\section{The radio data}\label{sec:radiodata}

In this work we make use of data from the NRAO VLA Sky Survey (Condon et al. 
\cite{Condon}). The NVSS started in 1993 with the VLA in D and DnC 
configurations and has recently been completed. The NVSS covers the whole sky 
north of $\delta = -40\degr$ at the frequency $1.4$~GHz with resolution 
$45\arcsec$. 

Data products consists of $2326$ $\approx 4\degr \times 4\degr$ maps in Stokes 
I, Q, and U with pixel size $15\arcsec$ and rms brightness fluctuations 
$0.45$~\bril. The positional rms in Right Ascension and Declination varies from
$<1\arcsec$ for relatively strong ($S > 15$~mJy) point sources to $7\arcsec$ 
for the faintest ($S = 2.3$~mJy) detectable sources. 

The positional accuracy, together with the low flux limit and moderate 
resolution of the survey makes the NVSS particularly suitable for the detection
of low-surface brightness extended structures and for the search of optical 
counterparts of radio sources.

A list of about $2 \times 10^6$ discrete sources is available as well, and has
been extracted from the survey images  by fitting elliptical Gaussians to all 
significant peaks (Condon et al. \cite{Condon}). In the compilation of this 
list, hereafter NVSS--NRAO catalogue, no attempt is made to classify sources 
according to their morphology (double or pointlike sources).

Nevertheless, when one wants to make optical identifications, a crucial point 
is the knowledge of the source structure. If a double source, for which we can 
expect to find the optical counterpart near the radio barycentre position, is 
erroneously treated as two single components, the identification procedure can 
lead to misleading results, thus seriously affecting the completeness and 
reliability of the identification program.

The blind use of a list of fitted components like the NVSS--NRAO catalogue is 
thus not optimal if one wants to get a radiogalaxy sample characterized by well
defined statistical properties, suitable for further astrophysical 
applications. For this reason, we developed our own algorithm for the
extraction of a radio source catalogue from the radio maps and for the 
morphological classification of the detected sources, as will be discussed in 
the next Sections.

\section{The radio source extraction algorithm}\label{sec:catextraction}

The operations performed by the extraction algorithm are divided in five
modules: the source detection, the 1-Gaussian fit module, the evaluation of 
fit reliability, the 2-Gaussian fit module and the classification of double 
sources. More details on the operations performed by the algorithm are given in
Appendix~\ref{sec:apa}.

\subsection{Source detection}\label{sec:sourdet}

The algorithm reads each NVSS FITS map, consisting in a $1024 \times 1034$ 
pixel matrix ($1 \mathrm{pixel} = 15\arcsec$), and then looks for emission 
peaks: we adopted a threshold flux of $S_\mathrm{P} = 2.5$~\bril, corresponding
to the $5\sigma$ level for the NVSS survey (rms noise on NVSS I images is
$\approx 0.45$~\bril, Condon et al. \cite{Condon}).
A different detection threshold has been applied to two sky regions where 
strong residual diffraction lobes due to the presence of a very bright 
($\sim 2.5$~Jy) and extended source are found. To avoid detecting a large 
number of spurious sources, for these regions we evaluated the local noise and 
selected only those peaks with $S_\mathrm{P} \ge 5 \sigma_\mathrm{local}$.

A submatrix of $15 \times 15$ pixels ($\sim 3.8\arcmin \times 3.8\arcmin$)
around each peak is built, defining the region over which the operations 
described in the next Sects. are executed.

\subsection{Fit with one Gaussian component}\label{sec:onefit}

A fit with a circular Gaussian function of fixed ${\rm FWHM} = 45\arcsec$ is 
performed over each submatrix (see Appendix~\ref{sec:apa}); the FWHM of the 
fitting function has been chosen to reproduce the nominal beam of the NVSS.

The use of a fixed FWHM has the consequence that it is not possible to 
determine the angular dimension -- and thus the integrated flux -- of the
radio sources. Nevertheless, some tests showed that the use of a Gaussian of 
variable size is not advantageous when fitting sources at low flux levels 
($\simlt 8 \sigma$), whose resulting positions and fluxes were found to be
inaccurate. We decided thus to fix the dimensions of the fit function and to 
perform a fit with two Gaussian components in those cases when the 
one-component fit was not satisfactory. In Sect.~\ref{sec:fitrel} the criteria 
for performing a 2-component fit are described.

Input parameters for the 1-component fit are $x$ and $y$ peak pixel coordinates
of the submatrix central point, and the measured flux at that pixel. For each 
source, the algorithm computes the fit rms $\Sigma_\mathrm{1fit}$ (see 
Appendix~\ref{sec:apa}), which is used as a discriminant for the execution of 
the 2-Gaussians fit.

\subsection{Fit reliability}\label{sec:fitrel}

Inspecting the results obtained from the 1-component fit for some test sources,
we found that they are not satisfactory in terms of positional and photometric 
accuracy when the fit rms 
$\Sigma_\mathrm{1fit} > 0.6$~mJy~pixel$^\mathrm{-1}$.
The distribution of $\Sigma_\mathrm{1fit}$ values in different flux bins showed
that the percentage of sources with $S_\mathrm{P} < 5.0$~\bril~ for which
$\Sigma_\mathrm{1fit} \ge 0.60$~mJy~pixel$^\mathrm{-1}$ is reasonably low, of 
the order of $10\%$. We thus decided to apply a 2-component fit only to those 
sources with both $\Sigma_\mathrm{1fit} > 0.60~{\rm mJy~pixel^{-1}}$ and
$S_\mathrm{P} \ge 5.0~{\rm mJy~beam^{-1}}$. If it happens that 
$\Sigma_\mathrm{2fit} > \Sigma_\mathrm{1fit}$ ($\approx 4\%$ of these sources),
then the 1-component solution is restored.

There are however two categories of sources for which the above criterion did 
not guarantee good results with the 1-component fit, and required a different 
approach: these cases are what we called ``extended'' and ``multiple'' sources.

When in presence of ``extended'' sources, whose flux distribution presents a 
``plateau'' instead of a well defined maximum, the algorithm can detect more 
than one emission peak, and attempts to perform as many fits: this happens to 
about the $6\%$ of the sources fitted with 1 Gaussian component, with no
dependence on their flux. It has been possible to identify two different 
situations: if the distance between the fitted positions is less than 
$4\arcsec$ then is always 
$\Sigma_\mathrm{1fit} < 0.60$~mJy~pixel$^\mathrm{-1}$. 
For distances between $4\arcsec$ and $45\arcsec$, on the contrary, at least one
fit has $\Sigma_\mathrm{1fit} > 0.60$~mJy~pixel$^\mathrm{-1}$.
In the first case we verified that $1$ Gaussian with fixed FWHM reproduces the 
source correctly: the 1-component fit is considered valid, by assigning to the
radio source the position of the barycentre of the multiple fits. 
In the second case, the effect of the source extension is not negligible and 
the highest values of $\Sigma_\mathrm{1fit}$ and $S_\mathrm{P}$ among those 
fitted are attributed to the source, which is thus forced to the 2-component 
fit.

A further class of sources, the ``multiple'' ones, has been identified during
the implementation of the 2-component fit module: the dimension of the fit 
submatrix is such that the number of times it contains two sources is not 
negligible. We found that the presence of more than one source in the same fit 
submatrix seriously affects the minimization process and the accuracy of 
fitted parameters.

We took into consideration these situations by introducing the following
criterion: all those sources having a neighbour inside $2.5\arcmin$, with at 
least one of them having $\Sigma_\mathrm{1fit} > 0.6$~mJy~pixel$^\mathrm{-1}$, 
are considered ``double''. In such a case, a new fit submatrix is defined 
around the central position between the two components and the source is 
forced to the 2-component fit. A distance of $2.5\arcmin$ guarantees that the 
structure of both components is well represented in the region defined by the 
fit submatrix.

To keep track of the different operations and adopted criteria, multiple and 
extended sources have been marked with control flags. A further analysis of the
classification of double sources has been made as the final step in the 
construction of the radio source catalogue (see 
Sect.~\ref{sec:radiodouble}).

\subsection{Fit with two Gaussian components}\label{sec:twofit}

The 2-component fit models sources with two Gaussian functions, each having
${\rm FWHM} = 45\arcsec$. Input parameters needed to describe the two functions
are: $x$ and $y$ peak pixel coordinates of the submatrix central point, peak 
flux, distance in $x$ and $y$ between the two components (in pixels from the 
barycentre), logarithmic ratio of fluxes of the two components.
Obviously, this amounts to assume that the brightness distribution of the 
source is modelled as the sum of two pointlike sources.

Even if for true double radio sources it is seldom found a flux ratio 
$S_\mathrm{1}/S_\mathrm{2} > 4$, we allowed this parameter to be as high as 
$10$, with a lower limit for the flux of a single component 
$S_\mathrm{P} = 1.5$~\bril. This choice proved to be useful to correctly fit 
the flux of those ``extended'' sources discussed in Sect.~\ref{sec:fitrel}.
In fact, due to our choice of fixed--size Gaussian functions, when dealing with
``extended'' sources for which a second peak is not detected, the extraction 
algorithm may need a second component to correctly fit the source flux.

For each double source the algorithm evaluates total flux and barycentre
position, as well as flux, coordinates and separation of the two components.
The fit rms $\Sigma_\mathrm{2fit}$ is computed similarly to what is done 
for the 1-component fit; if $\Sigma_\mathrm{2fit} \ge \Sigma_\mathrm{1fit}$, 
and depending on the source control flags (if there are any), the 1-component 
fit may have been considered valid.

\begin{figure}
\resizebox{\hsize}{!}{\includegraphics{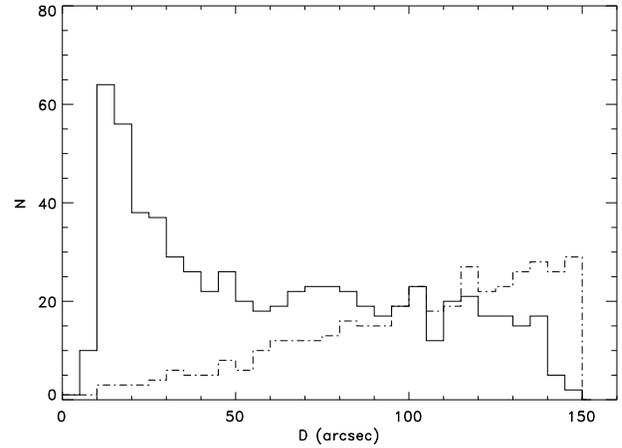}}
\caption{
Distribution of the separation between components, $D$, for $660$ double 
catalogue radio sources (solid line) and $409$ random doubles (dashed line), 
belonging to the $6$ maps we examined (see text).}
\label{fig1}
\end{figure}

\subsection{Classification of double radio sources}\label{sec:radiodouble}

The distinction made by the algorithm between single or double sources is
strongly influenced by the sky distribution of radio sources and by the 
characteristics of the fit procedure, so that a certain number of spurious 
associations classified as double on the basis of a positional coincidence of 
single, non interacting components is expected.
A further step in the compilation of the radio source catalogue is thus 
the estimate of the fraction of double radio sources that could have been so 
classified on the basis of the chance coincidence of two unrelated sources.

\begin{figure*}
\resizebox{\hsize}{!}{\includegraphics{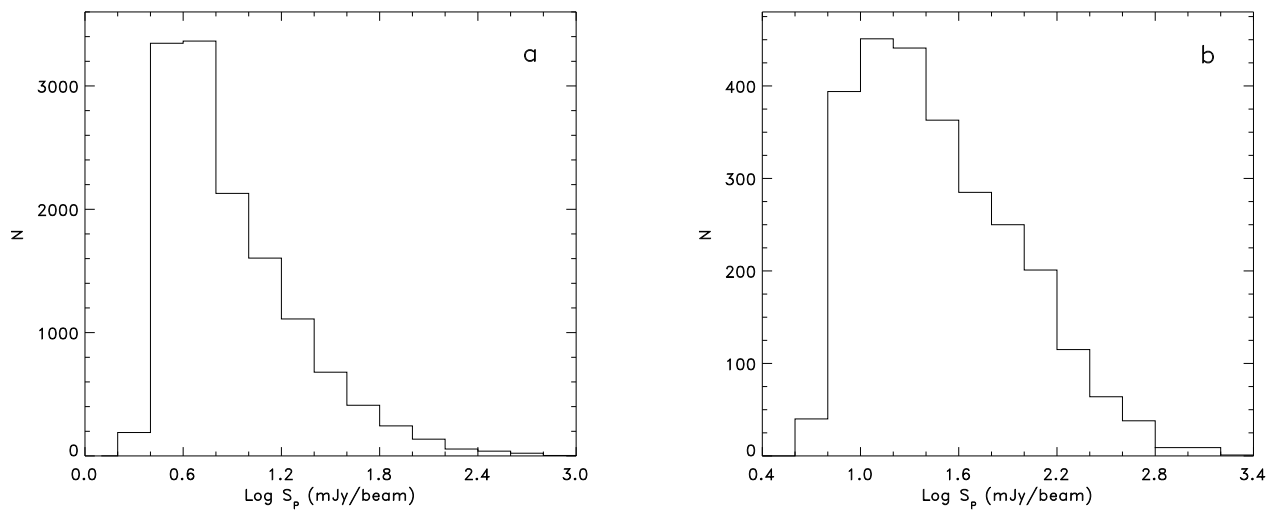}}
\caption{
Histogram of fluxes for pointlike ({\bf a}) and double ({\bf b}) radio sources
in the catalogue. There are $4$ pointlike sources with $Log S_\mathrm{P} >3.0$ 
and $1$ double source with $Log S_\mathrm{P} > 3.4$ not shown in these plots.
}
\label{fig2}
\end{figure*}

Given the observed surface density of NVSS radio sources, and under the
hypothesis that all of them are single sources, we looked at the probability
that a source has by chance a neighbour assuming a completely random sky 
distribution. We considered regions of $4\degr \times 4\degr$ belonging to $6$
NVSS maps we analyzed. For each region we generated $5$ random samples each 
containing as many positions as the detected NVSS sources (i.e. 
$1 \times n_\mathrm{single} + 2 \times n_\mathrm{doubles}$), associating to 
them values of the flux randomly chosen among the measured ones. 
We then looked for pairs in the random samples, i.e. sources having a neighbour
inside $2.5\arcmin$, that is the maximum distance we allowed for the 
classification of a double source.

We detected an average of $409$ spurious doubles over the $6$ maps we took into
consideration for this analysis. Over the same sky region, there are $660$ 
double radio sources in the catalog. In Fig.~\ref{fig1} the distributions
of the distance between components for catalogue double sources and random 
double sources are shown.

We distinguish three different contamination situations depending on the 
separation $D$ between the radio components. When $D \le 50\arcsec$ the mean 
contamination is about $13\%$ and we decided to keep as valid the 
algorithm classification: hereafter we will call these ``close'' double radio 
sources.
For separations larger than $100\arcsec$ the probability of chance coincidence 
is such that we can consider all of them as spurious doubles: the $709$ radio 
sources belonging to this interval have thus been included as two single 
components among the list of single sources.
In the interval between $50\arcsec$ and $100\arcsec$ a decision can hardly be 
made: the contamination rate for these radio sources is high ($\sim 61\%$) but 
the number of expected true doubles is not negligible. 
In order not to miss the corresponding optical identifications, we kept these 
sources (hereafter ``wide'' doubles) among the double ones, but for this group 
we followed a more careful procedure during the search of optical counterparts 
(see Sect.~\ref{sec:rgsample}).

Our estimate of the number of random doubles, derived under the assumption of a
uniform sky distribution of radio sources, does not take into account any 
effect due to clustering properties of radio sources.
However, there is indication that the clustering of radio sources on angular 
scales greater than the NVSS resolution is weak (Magliocchetti et al. 
\cite{Magliocchetti98}), and thus it should not significantly alter our 
results.

Due to our choice of a maximum separation between radio source components of 
$100\arcsec$, our catalogue of double radio sources does not include the class 
of ``giant'' doubles. For a reliable detection of such sources, additional 
radio data with a better angular resolution than that provided by the NVSS 
survey alone would be needed, to allow the determination of the morphological 
type and of the compact core component necessary for the identification of the 
optical counterpart.
Samples of giant radio sources have been selected on the basis of many
different criteria on their angular size, radio power and optical magnitude
(see e.g. Cotter et al  \cite{Cotter}; Machalski et al. \cite{Machalski}) 
so that it is not straightforward to give an estimate of the expected number of
missed giants in our catalogue.

\section{The radio source catalogue}\label{sec:radiocat}

The extraction algorithm has been applied to $31$ NVSS maps in the region of 
the South Galactic Pole. The algorithm classified as double $3371$ radio 
sources: among these, $709$ have distance between components $\ge 100\arcsec$ 
and have been included in the list of pointlike radio sources.

The resultant catalogue consists of $13\,340$ single sources and $2662$ double 
radio sources over $\approx 550$~sq.~degrees of sky. The distribution of peak 
fluxes is shown in Fig.~\ref{fig2}. The flux do not include any correction 
for the ``CLEAN bias'' ($\approx -0.3$~\bril~ for the NVSS): this has been 
taken into account when comparing our catalogue with the list of NVSS source 
components distributed by the NRAO (see Sect.~\ref{sec:testcat}).

\begin{figure*}
\resizebox{\hsize}{!}{\includegraphics{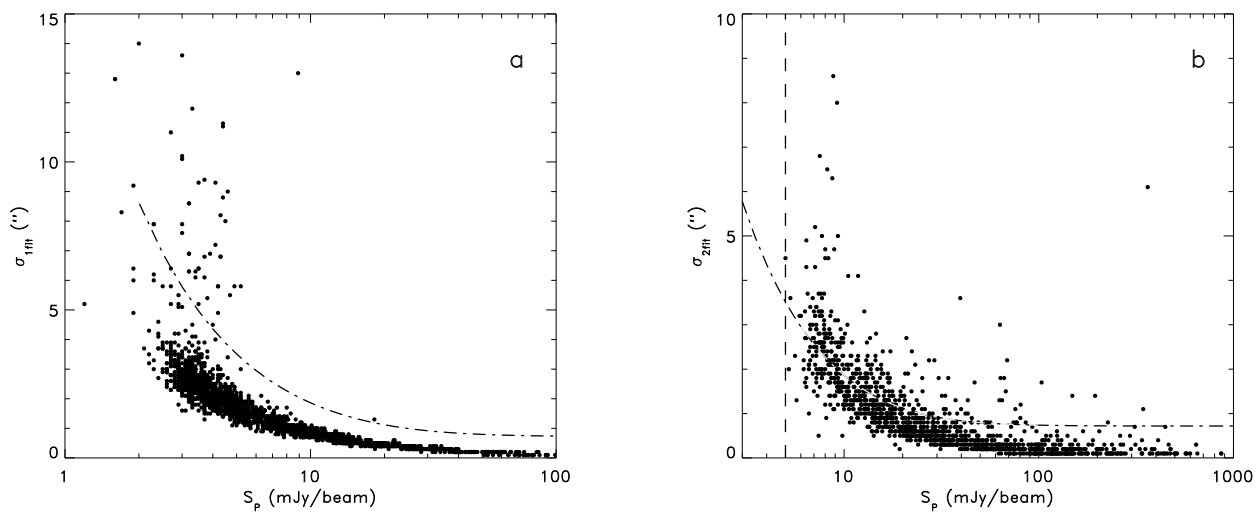}}
\caption{
Positional uncertainties estimated by the radio source extraction algorithm for
single ({\bf a}) and double ({\bf b}) radio sources as a function of peak flux.
The dot-dashed line represent the errors derived from the formulae in Condon 
et al. \cite{Condon}. In ({\bf b}) the dashed vertical line represents the flux
limit for the $2$ components fit (see Sect.~\ref{sec:twofit}).}
\label{fig3}
\end{figure*}

The catalogue is complete down to the NVSS flux limit $2.5$~\bril~ and, as can 
be seen in Fig.~\ref{fig3}, positional uncertainties estimated by the 
algorithm are in good agreement with those expected for NVSS sources (Condon 
et al. \cite{Condon}).

The electronic version of the radio source catalogue is available at the 
Centre de Donnees de Strasbourg (CDS).

\section{Tests on the catalogue}\label{sec:testcat}

As the accuracy in positions and fluxes can influence the photometric
completeness of the radio source catalogue as well as reliability and 
completeness of the optical identifications, before looking for radio source
counterparts some tests to verify the algorithm stability and reliability in 
computing positions and fluxes have been performed on the catalogue.

\begin{figure*}
\resizebox{\hsize}{!}{\includegraphics{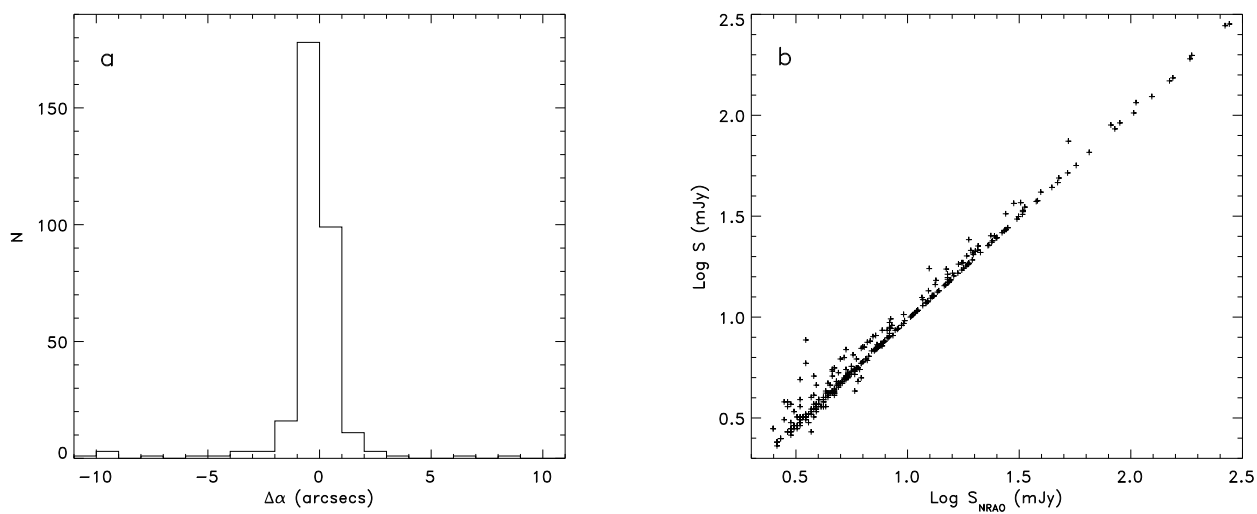}}
\caption{
{\bf a} Distribution of differences in Right Ascension for $323$ pointlike 
radio sources in our catalogue with respect to NVSS--NRAO catalogue positions.
{\bf b} NVSS--NRAO fluxes vs. radio source catalogue ones for the same 
set of sources. There are $4$ off-plot sources brighter than $160$~\bril.}
\label{fig4}
\end{figure*}

\subsection{Comparison with the NVSS--NRAO catalogue}\label{sec:catnrao}

A first test to asses the accuracy of fluxes and positions computed by the
radio source extraction algorithm described in the previous Sects. has been 
made with comparison to the NVSS--NRAO catalogue (Condon et al. \cite{Condon}).
The NVSS--NRAO catalogue does not provide any classification in single or
double radio sources, and simply gives a list of components fitted with an 
elliptical Gaussian of variable size: for this reason, we restricted this
quantitative analysis only to the single radio sources in our catalogue. 
A qualitative analysis of double radio sources has been made by visual 
inspection and described below.

We extracted from our catalogue a set of $323$ pointlike radio sources
belonging to the central $3 \times 3$ square degrees of the NVSS map I0016M24, 
and compared their positions and fluxes with those found in the NVSS--NRAO 
catalogue. To take into account the dependence of positional accuracy on the 
source flux, this analysis has been made in the three flux intervals
$S_\mathrm{P} \le 4$~\bril, $4 < S_\mathrm{P} \le 8$~\bril~ and 
$S_\mathrm{P} > 8$~\bril. The modules of the mean differences in Right 
Ascension and Declination were found to vary from $\approx 0.5$ to 
$\approx 0.02\arcsec$ with a dependence on flux, with rms varying from 
$\sim 2\arcsec$ for the lowest fluxes to $\sim 0.3\arcsec$ for sources brighter
than $8$~\bril.

Photometric accuracy has been tested by comparing peak fluxes in the NVSS--NRAO
catalogue with those determined by our extraction algorithm: these last result
to be on average slightly underestimated, the median of the differences being
$\Delta S_\mathrm{P} = -0.3$~\bril, of the order of the ``CLEAN bias'' term 
for which the published NVSS--NRAO flux values have been corrected.

In Fig.~\ref{fig4} the offset distribution in Right Ascension, similar to 
the one in Declination, and the difference in fluxes between our catalogue and 
the NVSS--NRAO one are shown for the $323$ considered sources.

During this analysis we found that in some cases a bright pointlike radio
source is split in more than one component by the NVSS--NRAO extraction 
algorithm.
This fact can be ascribed to the use of a totally automatic extraction
procedure, needed when managing such a huge amount of data. Nevertheless, it 
points out that the ``blind'' use of a component catalogue like the NVSS--NRAO 
one for optical identifications of radio sources can introduce contamination 
and incompleteness effects in the sample of of optical counterparts.

As the NVSS--NRAO catalogue does not make any attempt in classifying double
radio sources, a similar quantitative comparison has not been possible for
double systems and we limited our analysis to the visual inspection of $68$ 
cases of ``close'' and ``wide'' doubles in the catalogue. 
``Close'' pairs are generally fitted with $1$ component in the NVSS--NRAO 
catalogue and we found a good agreement between the NVSS--NRAO component 
position and the barycentre in the radio source catalogue.
A different situation exists for ``wide'' pairs in the radio source catalogue: 
their radio structure, as seen on the radio maps, is typical of classical 
double radio sources. In such cases, for which the NVSS--NRAO catalogue lists 
only the positions of the two components, the optical counterpart is clearly to
be searched near the radio {barycentre} position and thus would be missed if 
one makes a blind use of the NVSS--NRAO catalogue.

\subsection{Stability and reliability of the algorithm}\label{sec:catjmfit}

A further test has been made on the radio source catalogue by using sources in 
the overlapping regions between adjacent maps to check the stability of the 
extraction algorithm in reproducing fluxes and positions.

We compared catalogue data relative to $120$ single sources (half of them with 
fluxes larger than $10$~\bril) and $53$ double sources with those obtained by 
fitting the same sources with the AIPS task JMFIT.
We again found consistency with what predicted for NVSS sources: errors on the 
positions of single sources vary from $\approx 3\arcsec$ at fluxes lower than 
$5$~\bril, to $0.2\arcsec$ for $S_\mathrm{P} \ge 15$~\bril. For the double 
sources we find slightly larger values, $\sim 2\arcsec$ for 
$S_\mathrm{P} \ge 15$~\bril. 
We can conclude that both for pointlike and double radio sources in our 
catalogue, the positional accuracy is good enough to allow optical 
identifications with galaxies brighter than $b_\mathrm{J}=20.0$ down to the 
NVSS flux limit.

The variation of catalogue peak fluxes and peak fluxes obtained with JMFIT is 
$\simlt 1\%$ for single sources, similar to the values found examining radio 
sources in the overlapping regions; for double sources the fractional 
variation reaches the $\simgt 5\%$ for fluxes larger than $15$~\bril.
This higher value can be ascribed to a difficulty in representing the source 
with two Gaussians of fixed FWHM as the source flux increases. However, these 
results can be considered  satisfactory as the uncertainties are not such to 
compromise the reliability of the optical identifications.

\section{The optical identification procedure}\label{sec:optids}

The optical identification procedure has been applied separately to the three 
classes of NVSS radio sources in our catalogue: pointlike, ``close'' doubles 
and ``wide'' doubles. 
``Wide'' doubles are in fact affected by a non-negligible probability of being 
erroneously classified as double systems by our extraction algorithm, that is, 
we do not know a priori when the optical counterpart is to be expected 
near the radio barycentre, which we assume as the most likely position if the 
classification is correct (Venturi et al. \cite{Venturi}; Prandoni  et al. 
\cite{Prandoni}) or near the components. We thus have followed a careful 
approach in the identification of these sources, as will be detailed in 
Sect.~\ref{sec:rgsample}.

\begin{figure*}
\resizebox{\hsize}{!}{\includegraphics{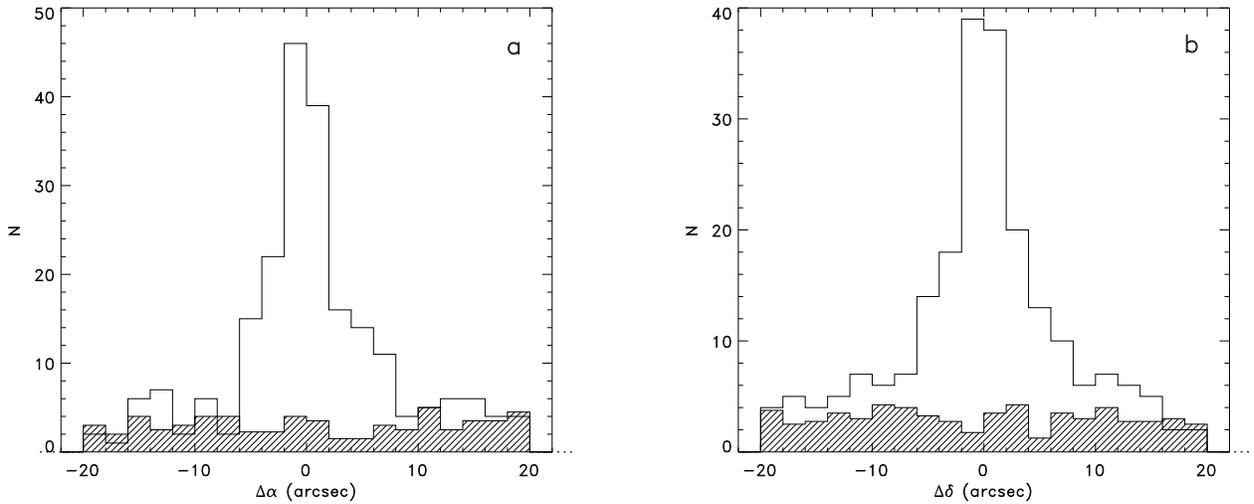}}
\caption{
Radio-optical offsets in Right Ascension ({\bf a}) and Declination ({\bf b})
for $218$ barycentres of NVSS radio sources in the catalogue (empty areas) and 
$62$ contaminants from random samples (shaded regions), optically identified 
with galaxies inside a box of size $40\arcsec$.}
\label{fig5}
\end{figure*}

``Close'' doubles have a low probability of being spurious associations of
single components, so that in principle they could be treated as the pointlike 
sources during the identification phase, by looking for counterparts near their
barycentres. Nevertheless, pointlike and ``close'' doubles have been kept
distinct in the identification phase, since we verified (see 
Sect.~\ref{sec:posuncert}) that for ``close'' doubles it is not possible to 
fulfil the requirements of the Likelihood Ratio method (De Ruiter et al. 
\cite{DeRuiter}).
This method -- to be applied in order to keep the contamination from 
radio-optical chance coincidence sufficiently low in the radiogalaxy sample -- 
has been used only for the list of optical counterparts of pointlike radio 
sources.

In the next Subsections we discuss the properties of the optical data used for 
the compilation of the radiogalaxy sample and the determination of 
radio-optical positional uncertainties, necessary to define the optimal radius 
for the search of optical counterparts. More details on the Likelihood Ratio 
method and on results from simulated samples are given in 
Appendix~\ref{sec:apb}.

\subsection{Optical data}\label{sec:optdata}

Optical identifications of radio sources in the NVSS catalog have been made 
with galaxies in the Edinburgh--Durham Southern Galaxy Catalogue (EDSGC, Nichol
et al. \cite{Nichol}). The EDSGC lists photographic $b_\mathrm{J}$ magnitudes 
for $\approx 1.5 \times 10^6$ galaxies over a contiguous area of 
$\sim 1200$~sq.~degrees at the South Galactic Pole.
For the construction of the catalogue glass copies of $60$ plates IIIa--J of 
the ESO/SERC Sky Survey at galactic latitude $|b_\mathrm{II}| \ge 20\degr$ have
been digitalized with the microdensitometer COSMOS (MacGillivray \& Stobie, 
\cite{MacGillivray}).
The automatic algorithm for star/galaxy classification implemented in COSMOS 
has been optimized so to achieve a completeness greater than $95\%$ and
stellar contamination less than $12\%$ for magnitudes $b_\mathrm{J} \le 20.0$.
Magnitudes have been calibrated via CCD sequences, providing a plate-to-plate
accuracy of $\Delta b_\mathrm{J}\simeq 0.1$ and an rms plate zero-point offset 
of $0.05$ magnitudes. 

The EDSGC catalogue becomes rapidly incomplete above $b_\mathrm{J}\simeq 20.5$,
thus we decided to consider only those galaxies with $b_\mathrm{J}\le 20.0$. 
With this conservative choice, the properties of the final identification 
sample (in terms of optical completeness and contamination) are well consistent
with the global ones of the EDSGC.

\subsection{Positional uncertainties and search radius}\label{sec:posuncert}

The search radius for optical identifications must be carefully chosen as it 
affects the completeness and reliability of the obtained radiogalaxy sample 
(see Appendix~\ref{sec:apb}). The optimal search radius is usually chosen on 
the basis of the total positional uncertainties, i.e. the combination of the 
radio error (comprehensive of the uncertainty introduced by the fit) which 
depends on source flux, and the accuracy on optical positions.

To avoid any kind of assumption on this term, we empirically determined the 
radio-optical positional accuracies from the distribution of the measured 
radio-optical offsets. Given the uncertainty in the classification of ``wide'' 
double radio sources, we made this analysis only for pointlike and ``close'' 
double sources.

We have first identified the $13\,340$ pointlike radio sources in the catalogue
with EDSGC galaxies brighter than magnitude $b_\mathrm{J} = 20.0$, looking for 
the nearest object inside a large square region of size $40\arcsec$, centered 
on each radio position.
The distributions of the observed radio-optical offsets in $\alpha$ and 
$\delta$ have been analyzed in different bins of radio flux, selected in order 
to contain approximately the same number of counterparts.

\begin{table}
\caption{Gaussian estimates of the total errors on coordinates, 
$\sigma_\mathrm{\alpha}$ and $\sigma_\mathrm{\delta}$, obtained from a fit on 
the $\Delta\alpha$ and  $\Delta\delta$ distributions of optical identifications
for catalogue pointlike radio sources. For each one of the $5$ considered flux 
intervals it is also given the mean number of contaminants per distance bin 
$C_\mathrm{med} /\mathrm{ bin}$, estimated by means of the $4$ control samples,
which represent the ``pedestal function'' over which the Gaussian distribution 
of the true identifications lies.
\label {tab:errorestim}
}
\begin{flushleft}
\begin{tabular}{c l c c c c c c} 
\hline\noalign{\smallskip}
& $S_\mathrm{P}$(~\bril) & & $C_\mathrm{med} / \mathrm{bin}$ & & $\sigma_\mathrm{\alpha}~{\rm (\arcsec)}$ & $\sigma_\mathrm{\delta}~{\rm (\arcsec)}$ &\\
\hline\noalign{\smallskip}
& $S_\mathrm{P} \le 3.5$ & & 3 & & 5.17 & 5.28 &\\
& $3.5 < S_\mathrm{P} \le 4.6$ & & 2 & & 4.11 & 5.76 &\\
& $4.6 < S_\mathrm{P} \le 7.5$ & & 3 & & 3.16 & 3.33 &\\
& $7.5 < S_\mathrm{P} \le 15.0$ & & 3 & & 2.04 & 2.27 &\\
& $S_\mathrm{P} > 15.0$ & & 3 & & 2.21 & 1.88 &\\
\hline
\end{tabular}
\end{flushleft}
\end{table}

Each offset distribution is the sum of two distinct distributions: a flat one 
due to the uniform distribution of spurious counterparts, plus a Gaussian one 
due to the true radio-optical associations.
To estimate the rms of this Gaussian, which is the desired positional
uncertainty, we first obtained an accurate measure of the mean level of
contaminants by making optical identifications of randomly generated samples.
We built $4$ control samples, each containing $13\,340$ random positions, and
looked for spurious optical counterparts in the same way as for catalogue 
sources.

By fitting the offset distributions with a Gaussian function plus a constant 
pedestal, given by the the contamination level in that flux range obtained from
the control samples, we evaluated the total positional uncertainties shown in 
Table~\ref{tab:errorestim}.

Similar values have been obtained repeating this analysis for the $1530$ 
``close'' double radio sources, i.e. looking for optical counterparts inside a 
box of size $40\arcsec$ centered on the radio barycentre, and using again
random samples to evaluate the contamination level.

On the basis of the estimated positional uncertainties, for the optical 
identification procedure of NVSS radio sources we thus adopted a search radius 
of $15\arcsec$.

The Likelihood Ratio method has subsequently been applied to the list of
counterparts of pointlike sources, to discard those cases that are 
statistically unlikely to be true radio-optical associations.
The same method proved to be inapplicable to the counterparts of ``close''
doubles, as in this case the hypothesis of Gaussian-distributed positional 
uncertainty, required by the Likelihood Ratio method, is not satisfied. As can 
be seen from Fig.~\ref{fig5}, in fact, an excess of true identifications is
found in the ``tails'' of the offset distributions.  These could correspond to
identifications of distorted radio sources, like Head--Tails, whose morphology 
is not completely resolved at the low NVSS resolution.

\section{The radiogalaxy sample}\label{sec:rgsample}

The list of NVSS pointlike radio sources identified with EDSGC galaxies 
brighter than $b_\mathrm{J} = 20.0$ inside a circle of radius $15\arcsec$ 
consists of $1061$ candidate counterparts, with an average of $254$ 
contaminants from the control samples: the initial contamination is thus 
$24\% \pm 2\%$. To this list of optical counterparts and to the counterparts 
found in the control samples we applied the modified Likelihood Ratio method 
described in Appendix~\ref{sec:apb}, evaluating $LR$ for each source using the 
positional uncertainty relative to its flux (see Table~\ref{tab:errorestim}).

The Likelihood Ratio cutoff value for rejecting a counterpart as unlikely to be
true was found to be $LR = LR_{\ast} = 1.9$. Out of the initial list of $1061$ 
candidates, the final sample of optical identifications of pointlike NVSS radio
sources thus consists of $926$ counterparts satisfying the condition 
$LR_{\ast} \ge 1.9$, while the number of contaminants in this sample, given by 
the mean number of spurious identifications in the control samples which have 
$LR \ge 1.9$, is $C_\mathrm{med}(LR_{\ast} \ge 1.9) \approx 145$.

The contamination percentage in the final sample of $926$ optical counterparts
of NVSS pointlike radio sources is thus $\sim 16\% \pm 1\%$, while due to the 
choice of the cutoff value for $LR$ we expect to lose $\sim 24$ true 
identifications. This corresponds to a completeness of $\simeq 97 \pm 1\%$ and 
to a reliability of $\simeq 84\% \pm 1\%$: we conclude that, with respect to 
the initial list of $1061$ candidate counterparts, the use of the modified 
Likelihood Ratio has sensibly lowered the contamination level without 
discarding a large number of real radio-optical associations. The 
identification percentage, expressed as the ratio between the number of true 
identifications and the total number of sources for which we looked for an
optical counterpart, is about 
$\theta_\mathrm{pointlike} = (926-145) / 13340 = 6\% \pm 0.2\%$.

\begin{figure*}
\vspace{13.6truecm}
\caption{
Optical identifications of double radio sources in the interval
$50\arcsec < D < 100\arcsec$: images are taken from the Digitized Sky Survey 
and contours represent radio emission in the NVSS. From left to right and from 
top to bottom: B00025, B01640, B01094, B02230, B02322, B02348, B00532, B00065, 
for which we identify the barycentre and one component with two different
galaxies; B00258, B00698, B01631 for which we identify the barycentre and $1$ 
component with the same galaxy and the second component with a different 
galaxy, and finally B01471 for which the barycentre and the two components are 
identified with $3$ different galaxies (see text).}
\label{fig6}
\end{figure*}

For the $1530$ ``close'' doubles we looked for an optical counterpart brighter 
than $b_\mathrm{J} = 20.0$ at a distance $\le 15\arcsec$ from the radio 
barycentre, finding $169$ identifications. The number of spurious 
identifications obtained from the control samples is $28 \pm 5$: the 
contamination percentage in the sample of optically identified ``close'' double
radio sources is then $16\% \pm 3\%$, while the identification percentage is 
$\theta_\mathrm{bar} = 9\% \pm 1\%$, consistent at the $3\sigma$ level with 
the value found for pointlike sources.

The identification procedure of the $1132$ ``wide'' double radio sources has
been made as follows: we first looked for an optical counterpart inside a 
radius of $15\arcsec$ both from the barycentre position and from the positions 
of the two components, and then inspected those cases where more than one 
identification is found for the same radio source.

We initially identified $232$ positions; in $156$ cases we identify either the
barycentre or one (or both) the components: in such cases, we consider valid 
the identification even if this does not mean that we are keeping the true 
optical counterpart.

In the remaining cases, a puzzling situation emerges as we find a counterpart
for both the barycentre and one component, or even for the barycentre and the
two components, so that a decision on the most reliable identification is 
difficult to make.
To discard or retain an identification, we decided to proceed as follows:
when we identify the barycentre and $1$ component with the same galaxy ($25$ 
cases), we assume that this happens because in the extraction algorithm we 
allowed high flux ratios.
In fact, when a common identification is found for the barycentre and for one 
of the components (normally the strongest) we consider valid the association 
with the barycentre, even if this is somewhat arbitrary.
Large values of $S_\mathrm{1} / S_\mathrm{2}$, up to $10$, are a feature 
introduced by our extraction algorithm and are not representative of the true 
distribution of flux ratios for double radio sources (see discussion in 
Sect.~\ref{sec:twofit}). To perform an analysis of flux ratios and arm--length 
ratios, and to compare it with other samples, we would need radio maps with 
much better resolution.

In the $12$ cases when we identify the barycentre and one component with two
different galaxies, or the barycentre and one component with the same galaxy 
but at the same time we find a different identification for the second
component, or finally we identify separately both the barycentre and the two 
components, we decided which is the most likely identification by visually
inspecting the field. These few cases are shown in Fig.~\ref{fig6}.

When the radio source structure is similar to a ``head--tail'' morphology, we 
considered the counterpart associated to the component corresponding to the 
radio source ``head''. In the presence of extended but symmetric radio 
morphologies we retain the identification in the barycentre. An example of 
spurious double can be represented by B1471 in Fig.~\ref{fig6}, the only 
case where we identify at the same time the barycentre and the two components 
with different galaxies. In this case we considered valid the two counterparts 
associated to the components of the double source.

By applying these criteria, we obtained a list of $193$ optical counterparts of
``wide'' double radio sources. Because of the presence of an ``intrinsic'' 
radio contamination, and given the subjective method adopted in the selection 
of true counterparts, it is possible to give only a lower limit to the 
contamination present in this list of identifications. The number of spurious 
identifications is 
$A \times N \times \rho_\mathrm{opt} = \pi \times (15)^2 \times (1132 \times 3)
\times 2.7 \times 10^{-5} = 65$, that is a contamination percentage of 
$\sim 28\%$.

A summary of the different contamination levels as well as the number of 
optical counterparts found for each radio morphology we are considering is
given in Table~\ref{tab:radiogal}.

\begin{figure*}
\resizebox{\hsize}{!}{\includegraphics{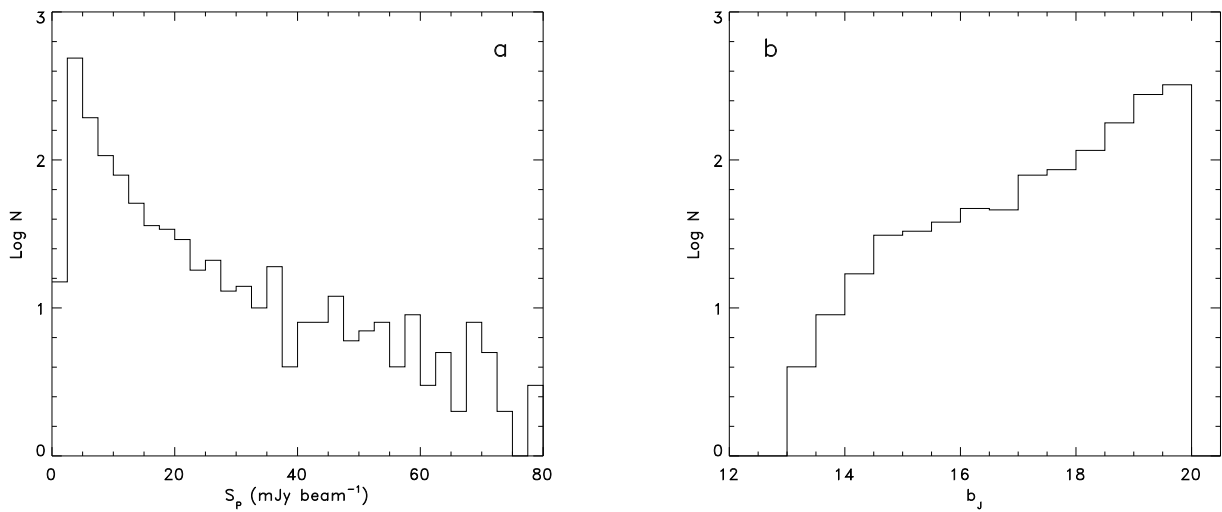}}
\caption{
Peak flux ({\bf a}) and magnitude ({\bf b}) distributions for the $1288$
NVSS radio sources optically identified with EDSGC galaxies. There are $67$ 
radiogalaxies brighter than $80$~\bril~ not shown in the flux histogram.}
\label{fig7}
\end{figure*}

The final radiogalaxy sample thus consists of $1288$ sources optically 
identified with galaxies brighter than $b_\mathrm{J} = 20$ in the EDSGC: in 
Fig.~\ref{fig7} the magnitude and flux distributions for the radiogalaxy 
sample are shown. The overall contamination in the radiogalaxy sample is 
$\sim 18\%$.
The identification percentage we find for NVSS sources is in agreement with 
what found by Magliocchetti \& Maddox (\cite{Magliocchetti}) for optical 
identifications of higher--resolution FIRST radio sources with APM galaxies in 
the equatorial region, once scaled to our radio flux and optical magnitude 
limits.
In Paper II we will use a subset of our radiogalaxy sample, characterized by a
higher reliability, to look for intermediate redshifts cluster candidates 
associated to NVSS radio sources.

\section{Summary}\label{sec:summary}

Aim of this work is to build a sample of radio-optically selected clusters of 
galaxies at
intermediate redshift in order to study the evolution and general properties of
groups and clusters as well as the effect of the environment on the radio 
emission phenomenon. In this paper we have discussed the compilation of a radio
source catalogue from $31$ NVSS radio maps covering the South Galactic Pole 
region, and the search of optical counterparts of these radio sources. The 
main reason to build a radio source catalogue alternative to the NVSS--NRAO 
publicly available one has been the need to classify radio sources according to
their morphology -- unresolved or double -- so to properly search for their 
optical counterparts.

Our radio source catalogue has been built by detecting emission peaks above the
detection threshold $S_\mathrm{P} \sim 2.5$~\bril~ and fitting Gaussian 
components with FWHM equal to the NVSS beam size to the selected peaks. The 
source detection algorithm first attempts a one-component fit to each peak and,
depending on the root mean square of the fit and on the distance between two 
neighbour peaks, if necessary a two-components fit is performed.

Classification of double radio sources has been done by first allowing the 
separation between components to be as large as $2.5\arcmin$ and compiling a 
first list of ``tentative'' double sources. Then, given the NVSS low 
resolution, a detailed analysis to discriminate between pointlike and double 
sources has been done by studying the probability of classifying two single, 
non interacting components as a double system on the basis of their separation.
From this analysis we found that the probability of two sources being a 
physically bound system is negligible when their distance is greater than 
$100\arcsec$.
These doubles have been removed from the ``tentative'' list and included as 
single components among the unresolved sources while, on the opposite, the 
classification of double sources is correct for those systems having 
$D \le 50\arcsec$ (``close'' doubles). 
In the intermediate range $50\arcsec < D < 100\arcsec$ the number of expected
spurious and true double sources are equivalent. These cases (``wide'' doubles)
have been included among double sources but for them a more careful optical 
identification procedure has been performed.

\begin{table}
\caption{The radiogalaxy sample: for each radio morphology class the number of
optical identifications and the contamination level are shown. The last row
gives these quantities for the whole sample.
\label {tab:radiogal}
}
\begin{flushleft}
\begin{tabular}{clccccccc} 
\hline\noalign{\smallskip}
& \hspace{0.1cm}Radio & & & N & & & Contamination &\\
& \hspace{0.1cm}Morphology & & & & & & & \\
\hline\noalign{\smallskip}
& Pointlike & & & 926 & & & 16 $\pm$ 1$\%$ &\\
& ``Close'' double & & & 169 & & & 16\% $\pm$ 3$\%$ &\\
& ``Wide'' double & & & 193 & & & 28$\%$ &\\
\hline\noalign{\smallskip}
& Total & & & 1288 & & & 18 $\%$ &\\
\hline
\end{tabular}
\end{flushleft}
\end{table}

The final radio source catalogue consists of $13\,340$ single and $2662$ 
double radio sources over $\approx 550$~sq.~degrees of sky, and is complete 
down to $S_\mathrm{P} = 2.5$~\bril.

A quantitative test to assess the accuracy of the radio source extraction
algorithm has been made comparing fluxes and positions of a set of radio 
sources in our catalogue with the correspondent values in the NVSS--NRAO 
catalogue. 
Since the NVSS--NRAO catalogue does not classify double radio sources this 
analysis has been possible for pointlike sources only.
We found that our results are in agreement with the ones in the NVSS--NRAO 
catalogue, well inside the predicted errors for the NVSS (Condon et al. 
\cite{Condon}). 
For what concerns double radio sources, we made a qualitative analysis by
visually inspecting a set of ``close'' and ``wide'' doubles and looking at 
their characteristics in the NVSS--NRAO catalogue. We found a good agreement in
flux and positions for ``close'' doubles, while in most cases ``wide'' ones 
clearly show a classical double radio morphology on the maps.
In such cases, where the optical counterpart should be looked near the radio 
barycentre, the use of the NVSS--NRAO catalogue without a re-processing to 
detect double sources, would result in a loss of optical identifications and 
thus in a less complete radiogalaxy sample.

Optical identifications of radio sources in our catalogue have been made with 
EDSGC galaxies (Nichol et al. \cite{Nichol}) down to a limiting magnitude of 
$b_\mathrm{J} = 20.0$ and adopting a search radius of $15\arcsec$.

Different strategies have been applied for the search of optical counterparts
of pointlike or double radio sources. For the latter, the probability of having
classified as a double system two physically disjointed sources on the basis of
their superposition in the sky is in fact dependent on the distance between the
two components.
The optical identification of the $13\,340$ pointlike radio sources led to a 
sample of $926$ radiogalaxies. The statistical completeness and reliability of 
this sample have been evaluated by means of the modified Likelihood Ratio 
method proposed by De Ruiter et al. (\cite{DeRuiter}) (see 
Appendix~\ref{sec:apb}), to properly take into account  the true optical 
surface distribution of galaxies in the sky. 
This sample is complete to $97\% \pm 1\%$ and reliable to $84\% \pm 1\%$, with 
an identification percentage of $6\% \pm 0.2\%$.

The optical identification of $1530$ ``close'' double radio sources (distance 
between components $D \le 50\arcsec$) has been made looking for a counterpart 
near the barycentre position. For these sources, the probability of being a 
spurious double is low, $\simlt 13\%$. We optically identified $169$ 
barycentres of ``close'' doubles; in this case it was not possible to apply the
modified Likelihood Ratio method to evaluate the reliability and completeness 
of the sample.
An estimate of the contamination level has been computed as the probability of 
chance radio-optical superposition on the basis of the average observed optical
surface galaxy density. We found a contamination of $16\% \pm 3\%$ for optical 
identifications of ``close'' double radio sources.

Optical identifications of ``wide'' doubles (distance between components 
$50\arcsec < D < 100\arcsec$) are made difficult by the high percentage of 
expected radio misclassification: the number of true radio associations is in 
fact comparable with the number of radio contaminants. We thus looked for 
optical counterparts both near the radio barycentre and near the radio 
components positions, visually inspecting those cases where more than one 
optical identification is found for the same radio source.
We found a list of $193$ optical counterparts of ``wide'' double radio sources,
with a contamination of the order of $\sim 28\%$: this contamination level must
be seen as a lower limit, as it does not take into account the joint 
probability of having a optical spurious identification near the barycentre of 
a spurious double radio source.

The final sample thus lists $1288$ radiogalaxies and represents a valuable
opportunity for the study of the multi-wavelength properties of the radiogalaxy
populations down to a low flux level.

This sample has been used to look for galaxy clusters associated to NVSS 
radiogalaxies: in a following paper (Zanichelli et al. \cite{Zanichelli}) we 
discuss the cluster selection strategy and the first observational results, 
that prove this technique to be a powerful tool for the selection of galaxy 
groups and clusters at intermediate redshift.

\appendix
\section{the Gaussian Fitting Algorithm}\label{sec:apa}

To define flux and accurate positions of radio sources from NVSS maps, we 
developed a code which performs a Gaussian bidimensional fit by means of a 
minimization process. Starting from $M$ functions in $N$ variables, 
$f_k(x_\mathrm{1},x_\mathrm{2},...,x_\mathrm{N})$, the routine MINSQ 
(Pomentale, \cite{Pomentale}) minimizes the sum:

\begin{eqnarray}
\phi^2(x_\mathrm{1},x_\mathrm{2},...,x_\mathrm{N}) = 
\sum _{k=1}^{M}\biggl\{f_{k}(x_\mathrm{1},x_\mathrm{2},
...,x_\mathrm{N})\biggr\}^2
\label{equ:minimize}
\end{eqnarray}

\noindent
where $M \ge N \ge 2$. The minimization process is iterated until the
difference between the function before and after the minimization is lower than
a user-selected value (``stopping rule''), or until a pre-defined maximum 
number of iteration is reached.
Each source to be fitted is represented with a circular Gaussian of 
${\rm FWHM} = \sigma$ and peak amplitude $A$:

\begin{eqnarray}
G(x,y) = A ~e^{- {x^2 +y^2 \over 2\sigma^2}}
\label{equ:gauss} 
\end{eqnarray}

\noindent
If the source image is composed of $M$ independent measures of the amplitude 
$a_k$, each one with a known associated error $\sigma$, the $f_k$ can be 
defined as:

\begin{eqnarray}
f_k = {[a_k - G(x_k, y_k)]^2 \over \sigma^2}
\end{eqnarray}

\noindent
Inserting this expression for the $f_k$ in the (\ref{equ:minimize}), the 
maximum-likelihood fit would be the one which minimizes the $\phi^2$.
In our case, the errors on the individual measurements are not a priori known: 
as a first approximation we could assume that they are constant over the image 
and equal to the mean survey rms ($\approx 0.45$~\bril), but this assumption
fails in presence of bright sources. We thus expressed the $f_k$ functions 
simply as the unweighted quadratic differences between the data and the fit at 
each pixel:

\begin{eqnarray}
\phi^2 = {\sum\limits_{k=1}^M f_k} = {\sum\limits_{k=1}^M 
(a_k - G(x_k,y_k))^2 }
\end{eqnarray}

\noindent
The value of $\phi^{2}_\mathrm{min}$ obtained from the minimization procedure 
and normalized to the number of functions $M$ is the estimated error associated
to the fit procedure. This uncertainty can be expressed as the sum in 
quadrature of a constant term, dependent on the map noise, plus a term 
proportional to the source flux through an a priori unknown constant:

\begin{eqnarray}
FF = \sqrt {\epsilon ^2 + (c \times S_\mathrm{P})^2}
\label{equ:estim}
\end{eqnarray}

\noindent
Thus, FF is not a good indicator of fit reliability due to its dependence on
source flux. To correct for this dependence, we determined $c$ as follows: 
first, we evaluated $\epsilon$ by analysing the distribution of FF for faint 
sources, for which the flux term in (\ref{equ:estim}) is negligible and the 
median value of the distribution of FF is a good approximation for $\epsilon$.
Second, introducing this value of $\epsilon$ in (\ref{equ:estim}) and 
considering bright sources, the value for the constant $c$ can be determined.

The fit uncertainty associated to each source is thus writable as:

\begin{eqnarray}
\Sigma = \sqrt {FF^2 - (c \times S_\mathrm{P})^2}
\end{eqnarray}

\noindent
and is evaluated both for 1-component and for 2-components fit.

Starting from source pixel coordinates, Right Ascension and Declination have
been computed by means of the conversion formulae for the sine projection
used in the NVSS:

\begin{eqnarray}
\alpha = \alpha_\mathrm{0} + \arctan\left ( x \over \cos \delta_\mathrm{0} 
\sqrt{1-x^2 -y^2}
- y \sin \delta_\mathrm{0} \right)
\label{eqn:alpha} 
\end{eqnarray}

\begin{eqnarray}
\delta = \arcsin\left (y \cos \delta_\mathrm{0} + \sin \delta_\mathrm{0} 
\sqrt{1-x^2 -y^2}
\right )
\label{eqn:delta} 
\end{eqnarray}

\noindent
where ($\alpha_\mathrm{0} ,\delta_\mathrm{0}$) are the central Right Ascension 
and Declination of the map, and ($\alpha ,\delta$) are those of a source with 
known pixel coordinates ($x,y$).

\section{The modified likelihood ratio -- using control samples}\label{sec:apb}

The sample resulting from an optical identification program is characterized by
a contamination level, which depends on the number of spurious identifications,
and a completeness level, which is the percentage of true radio-optical 
associations we were able to correctly identify on the basis of the chosen 
search radius.

We have a ``correct'' identification when the combined radio and optical 
positional uncertainties are such that the true counterpart of the radio 
source, if it exists, does not lie outside the area defined by the search 
radius and, at the same time, the first (nearest) contaminant is not closer to 
the radio source than the identification itself.

In the case when a correct identification does not exist (empty field), we will
misidentify as true a contaminant each time a galaxy is found inside the search
region. The percentage of identification is defined as the fraction of correct 
identifications with respect to the total number of radio sources for which an 
optical counterpart has been looked for.

The completeness of an optical identification program represents the fraction 
of correct identifications among the radio sources having an optical 
counterpart, while the reliability is defined as the fraction of counterparts 
that are true radio-optical associations, i.e. it is the complement to the 
contamination level in the sample.

Under the hypothesis that the positions of a radio source and its optical
counterpart are intrinsically coincident, it is possible to define the {\it a 
priori} probability $p(r\mid id)$ that the radio-optical offset is found in the
distance interval ($r, r+dr$) due to the positional uncertainties.
Similarly, under the hypothesis that the counterpart is a contaminant, it is
possible to define the a priori probability $p(r \mid c)$ that the 
contaminant is found inside ($r, r+dr$).

For each radio source it is then possible to define the Likelihood Ratio $LR$
as the ratio between these two probabilities: an optical counterpart is 
considered as the true radio-optical association if $p(r\mid id)$ is greater 
than $p(r \mid c)$ by a factor $LR_{\ast}$ to be determined.

Nevertheless, what is actually computable from an identification program are 
the a posteriori probabilities $p(id \mid r)$ and $p(c \mid r)$ that, 
having found a counterpart at a given distance $r$ from the radio source, we 
are dealing with the true identification or with a contaminant.

The Likelihood Ratio method (De Ruiter et al. \cite{DeRuiter}) makes use of 
the Bayes theorem to express $p(id \mid r)$ and $p(c \mid r)$ in terms of $LR$,
that is by means of the correspondent a priori probabilities 
$p(r \mid id)$ and $p(r \mid c)$:

\begin{eqnarray}
p(id|r) = p(id) \times p(r|id) / p(r)\\
p(c|r) = p(c) \times p(r|c) / p(r)
\label{eqn:pidpc1}
\end{eqnarray}

\noindent
where $p(r)$ is the probability to find an object (irrespective if a
contaminant or the true identification) at a distance between $r$ and $r+dr$ 
from the radio source; $p(id)$ is the {\it a priori} probability to find the 
optical counterpart of a radio source and $p(c) = 1 - p(id)$ the probability to
find a spurious identification.

By applying the Bayes theorem and under the assumption that the true 
identification is always the nearest object to the radio source, $p(id \mid r)$
and $p(c \mid r)$ can be written as:

\begin{eqnarray}
p(id|r) = {\vartheta LR(r) \over {\vartheta LR(r)+1}}\\
p(c|r) = {{1} \over {\vartheta LR(r)+1}}
\label{eqn:pidpc2}
\end{eqnarray}

\noindent
where $\vartheta = \theta / (1 - \theta) $ and $\theta$ is the a priori unknown
percentage of expected true identifications. The latter can be estimated as the
sum of the probabilities for each individual identification to be real, 
normalized to the total number of counterparts found.
The quantities $\theta$ and $p(id \mid r)$ are not independent and the solution
for $\theta$ is found iteratively. The total number of expected true 
identifications, $N_\mathrm{id}$, is given by 
$N_\mathrm{id} = \theta N_\mathrm{tot}$, where $N_\mathrm{tot}$ is the total 
number of radio sources for which an optical counterpart is searched. 
Once $\theta$ is determined, the reliability and completeness of the final 
identification sample can be defined as a function of the cutoff value 
$LR_{\ast}$:

\begin{eqnarray}
C = 1 - \sum\limits_{LR_i < L} p_i(id|r) / N_\mathrm{id}
\label{eqn:compLR}
\end{eqnarray}

\begin{eqnarray}
R =1 - \sum\limits_{LR_i \ge L} p_i(c|r) / N( LR > L)
\label{eqn:relLR}
\end{eqnarray}

\noindent
Where $N(LR > L)$ is the total number of identifications having $LR > L$. The 
value for $LR_{\ast}$ is determined by studying the behaviour of $C$ and $R$ as
a function of $LR$, finding the value of $LR$ that maximizes $(C+R)/2$.

In general, the value of $LR_{\ast}$ is close to $\sim 2.0$, that means to 
consider true all those identifications for which the a priori probability of 
having correctly identified the radio source is twice the a priori probability 
of having a contaminant.

One critical factor in the Likelihood Ratio method proposed by De Ruiter et al.
(\cite{DeRuiter}) is the assumption of a constant optical surface density of 
galaxies. 
This does not allow to keep into account the real galaxy clustering and thus 
can heavily affect the estimates of $C$ and $R$.
To avoid this limitation, we applied a modified version of this method, which 
makes use of control samples to properly evaluate the contamination level in 
the optical identification samples.

Control samples of the same size as the radio source catalogue are built by 
assigning to each entry a random position and, once defined the radius of the 
search region, optically identified with galaxies as is done for the 
radiogalaxy sample. We can write the expected number of contaminants in the 
final identification sample as the average of the spurious identifications 
found in each control sample: $C_\mathrm{med}$.
The expected number of true identifications will thus be given by the
difference between the total number of counterparts found, $N$, and the mean 
number of contaminants: $N_\mathrm{id} = N - C_\mathrm{med}$. We can obtain 
also the identification percentage $\theta = N_\mathrm{id} / N_\mathrm{tot}$, 
where $N_\mathrm{tot}$ is the total number of radio sources for which we 
have searched an optical counterpart.

According to the Likelihood Ratio method, the completeness expresses the 
fraction of real identifications for which $LR\ge L$, so we can write:

\begin{eqnarray}
C = 1 - (N(LR < L) - C_\mathrm{med}(LR <L)) /N_\mathrm{id}
\label{eqn:comLRmod}
\end{eqnarray}

\noindent
The term in parenthesis is the number of true identifications (i.e. excluding 
the contaminants) that are lost due to the choice of the cutoff value 
$LR_{\ast}$.

Similarly, we can write for the reliability: 

\begin{eqnarray}
R = 1- (C_\mathrm{med}(LR \ge L))/ N(LR\ge L)
\label{eqn:relLRmod}
\end{eqnarray}

\noindent
That is, $R$ is defined in terms of the fraction of contaminants that are
included in the sample due to the choice of the cutoff value $LR_{\ast}$.


\begin{thebibliography}{}

\bibitem[1989]{Abell}
Abell G.O., Corwin H.G., Olowin R.P., 1989, ApJS 70, 1

\bibitem[1993]{Allington--Smith}
Allington--Smith J.R., Ellis R.S., Zirbel E.L., Oemler A., 1993, ApJ 404, 521

\bibitem[1988]{Baum}
Baum S.A., Heckman T., Bridle A., van Breugel W., Miley G., 1988, ApJS 68, 643

\bibitem[1995]{Becker}
Becker R.H., White R.L., Helfand D.J., 1995, ApJ 450, 559

\bibitem[1984]{Bicknell84}
Bicknell G.V., 1984, ApJ 286, 68

\bibitem[1986]{Bicknell86}
Bicknell G.V., 1986, ApJ 300, 591

\bibitem[2000]{Blanton}
Blanton E.L., Gregg M.D., Helfand D.J., Becker R.H., White R.L., 2000, ApJ 531,
118

\bibitem[1984]{Bridle}
Bridle A.H., Perley R.A., 1984, ARA\&A 22, 319

\bibitem[1994]{Burns}
Burns J.O., Rhee G., Owen F.N., Pinkney J., 1994, ApJ 423, 94

\bibitem[1984]{Butcher}
Butcher H.R., Oemler A., 1984, ApJ 285, 426

\bibitem[1998]{Condon}
Condon J.J., Cotton W.D., Greisen E.W., et al., 1998, AJ 115, 1693

\bibitem[1996]{Cotter}
Cotter G., Rawlings S., Saunders R., 1996, MNRAS 281, 1081

\bibitem[1994]{Dalton}
Dalton G.B., Efstathiou G., Maddox S.J., Sutherland W.I., 1994, MNRAS 269, 151

\bibitem[1977]{DeRuiter}
De Ruiter H.R., Willis A.G., Arp H.C., 1977, A\&AS 28, 211

\bibitem[1982]{Feigelson}
Feigelson E.D., Maccacaro T., Zamorani G., 1982, ApJ 255, 392

\bibitem[1990]{Gioia}
Gioia I.M., Henry J.P., Maccacaro T., et al., 1990, ApJ 356, L35

\bibitem[1992]{Henry}
Henry J.P., Gioia I.M., Maccacaro T., Morris S.L., Stocke J.T., 1992,
ApJ 386, 408

\bibitem[1991]{Hill}
Hill G.J., Lilly S.J., 1991, ApJ 367, 1

\bibitem[1996]{Ledlow}
Ledlow M.J., Owen F.N., 1996, AJ 112, 9

\bibitem[1992]{Lumsden}
Lumsden S.L., Nichol R.C., Collins C.A., Guzzo L., 1992, MNRAS 258, 1

\bibitem[1984]{MacGillivray}
MacGillivray H.T., Stobie R.S., 1984, Vistas Astr. 27, 433

\bibitem[2001]{Machalski}
Machalski J., Jamrozy M., Zola S., 2001 A\&A 371, 445

\bibitem[1998]{Magliocchetti98}
Magliocchetti M., Maddox S.J., Lahav O., Wall J.W., 1998, MNRAS 300, 257

\bibitem[2001]{Magliocchetti}
Magliocchetti M., Maddox S.J. 2001 astro-ph/0106429

\bibitem[1999]{Miller}
Miller N.A., Owen F.N., Burns J.O., Ledlow M.J., Voges W., 1999, AJ 118, 1988

\bibitem[2000]{Nichol}
Nichol R.C., Collins C.A., Lumsden S.L., 2000, submitted to ApJS

\bibitem[1968]{Pomentale}
Pomentale T., 1968, CERN Computer Centre, Program Library

\bibitem[1996]{Postman}
Postman M., Lubin L.M., Gunn J.E., et al., 1996, AJ 111, 615

\bibitem[2001]{Prandoni}
Prandoni I., Gregorini L., Parma P., et al., 2001, A\&A 369, 787

\bibitem[1988]{Prestage}
Prestage R.M., Peacock J.A., 1988, MNRAS 230, 131

\bibitem[1998]{Rosati}
Rosati P., della Ceca R., Norman C., Giacconi R., 1998, ApJ 492, L21

\bibitem[1999]{Scodeggio}
Scodeggio M., Olsen L.F., da Costa L., et al., 1999, A\&A 137, 83

\bibitem[1997]{van Haarlem}
van Haarlem M.P., Frenk C.S., White S.D.M., 1997, MNRAS 287, 817

\bibitem[1997]{Venturi}
Venturi T., Bardelli S., Morganti R., Hunstead R.W., 1997, MNRAS 285, 898

\bibitem[2001]{Zanichelli}
Zanichelli A., Scaramella R., Vettolani G., et al., 2001, 
A\&A in press, Paper II

\bibitem[1989]{Zhao}
Zhao J.H., Burns J.O., Owen F.N., 1989, AJ 98, 64

\bibitem[1996]{Zirbel96}
Zirbel E.L., 1996, ApJ 473, 713

\bibitem[1997]{Zirbel97}
Zirbel E.L., 1997, ApJ 476, 489


\end{thebibliography}
\end{document}